\documentclass[12pt]{iopart}
\usepackage{epsfig,color,multirow}
\begin{document}

\title[Gated combo nanodevice for sequential operations on single electron spin]{Gated combo nanodevice for sequential operations on single electron spin}

\author{S. Bednarek and B. Szafran}
\address{Faculty of Physics and Applied
Computer Science, AGH University of Science and Technology, al.
Mickiewicza 30, 30-059 Krak\'ow, Poland}

\begin{abstract}

An idea for a nanodevice in which an arbitrary sequence of three
basic quantum single qubit gates - negation, Hadamard and phase
shift - can be performed on a single electron spin. The spin state
is manipulated using the spin-orbit coupling and the electron
trajectory is controlled by the electron wave function self-focusing
mechanism due to the electron interaction with the charge induced on
metal gates. We present results of simulations based on iterative
solution of the time dependent Schr\"odinger equation in which the
subsequent operations on the electron spin can be followed and
controlled. Description of the moving electron wave packet requires
evaluation of the electric field within the entire nanodevice in
each time step.
\end{abstract}

\pacs{73.21.La, 73.63.Nm, 03.67.Lx}

\maketitle

\section{Introduction}
The power which is in principle offered by the idea of the quantum
computation stimulated an extensive effort toward realization of
quantum gates. Several quantum systems are studied for application
in encoding a quantum bit of information. The spin of an electron
confined inside a semiconductor nanostructure is considered
particularly promising \cite{spin1,spin2,spin3}. A number of
nanodevices were proposed and experimentally studied for operations
on the electron spin \cite{r1,r2,r3,r4,r5}. For practical
applications most useful are devices that function without microwave
radiation and without external magnetic fields. An example of such a
construction is the nanodevice presented in Ref. \cite{r5} in which the
operations on the electron spin are performed using the spin-orbit
coupling, and the transitions between the spin states are induced by
AC voltages applied to the electrodes. The spin-orbit coupling can
also be used for operations on the electron spins in opened quantum
rings \cite{p1,p2}. In Ref. \cite{p1} a system of quantum rings is
proposed for spin transformations according to three basic quantum
gates: NOT, phase shift and Hadamard transform in a spin-dependent scattering
experiment. In Ref. \cite{p2} a programable array of rings was
proposed to split the incoming electron wave function is split into two parts
each corresponding to an
opposite spin orientation. In these papers \cite{p1,p2} time independent
scattering problem is solved, with the electron spread all over the space
with position dependent spin orientation.
 Recently, we proposed \cite{a2}
a different device in which the position of the electron is localized in form
of a wave packet which is self-focused by the interaction of the metal with the
charge induced on the surface of metal electrodes which are deposited on top
of the nanostructure and form induced quantum wires or induced quantum dots \cite{a1}.
The electron localized in the wave packet has a definite spin orientation in every moment
in time, and operation on the spin are performed by electron motion within the nanostructure.
In contrast to the spin-dependent scattering approaches in open structures,
the operations on the electron spin are performed in a compact nanodevice
and result of time-evolution operators
according to the original theory of quantum computation.
In the proposed device
\cite{a1} one can set the electron in motion along precise
trajectories. Due to the interaction of the electron with the charge
induced on the metal electrodes the electron wave function becomes
self-focused \cite{a3,a4} forming a stable wave packet that can be
directed to a chosen location within the nanostructure with
probability 1.

Systems proposed in Ref. \cite{a2} exploit the Rashba coupling
\cite{ras} in the absence of the Dresselhaus \cite{dres} coupling.
They can be produced in semiconductors of the diamond structure, in
Si or Ge for instance. A similar device working in III-V or II-VI
material of zinc blende structure will work in presence of the
Dresselhaus coupling resulting of inversion asymmetry of the
crystal lattice. For both Rashba and Dresselhaus coupling present
and $y$ direction chosen in the growth direction the spin rotations with respect to the $x$ and the $z$
axes will occur for the electron moving in spatial directions within a plane of confinement $(x,z)$
forming a narrow angle whose specific value depends on the
the coupling constants. A practical realization of such a
device is possible although more challenging then the ones
in which only one type of the spin-orbit coupling is present.
In the present paper
we assume that the quantum well in which the electron moves
is perfectly symmetric so that the Rashba coupling constant is
zero. Accordingly, we will consider the Dresselhaus coupling only.
Then, the spin rotations around both the axes will be performed by
electron moving in perpendicular spatial directions.

\begin{figure}[ht!]
\centerline{\hbox{\epsfysize=100mm
               \epsfbox[67 355 522 820] {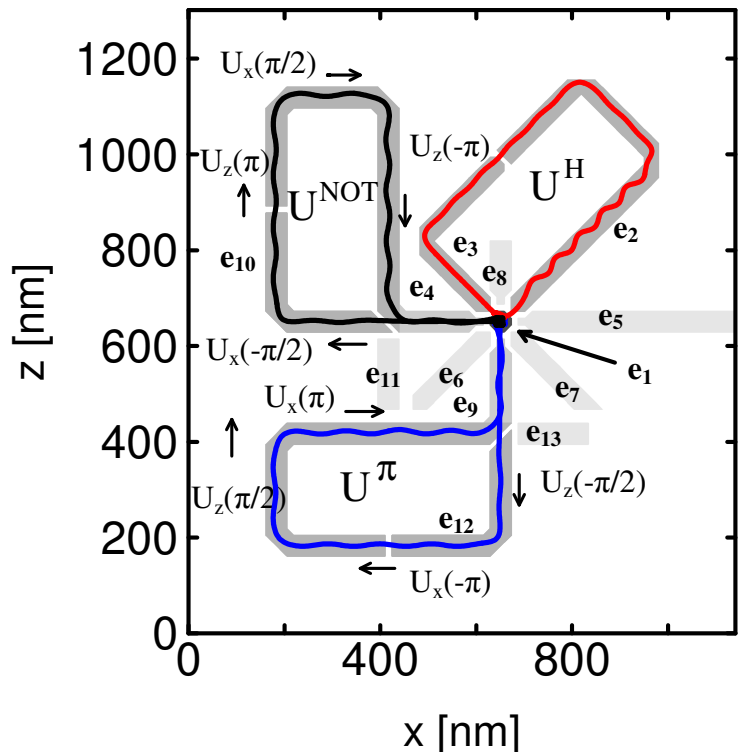}\hfill}}
\caption{System of thirteen electrodes deposited on top
of the planar semiconductor structure. The electrodes $e_1,e_2\dots,e_{13}$
are plotted in grey. A quantum dot induced under the electrode $e_1$
is used as the starting and the ending point of each trajectory. Trajectories of the
electron at which its spin is subject to the negation $U^{NOT}$, phase shift $U^\pi$ and
Hadamard transform $U^H$ are marked by black, blue and red curves, respectively.
\label{f2}}
\end{figure}

\section{Theory}
Systems discussed below are based on a planar semiconductor structure
as previously considered in Refs. \cite{a1,a2}. It contains a 10 nm
wide quantum well sandwiched between tunnel barriers each 10 nm thick.
The structure is separated of the strongly doped substrate by a
undoped buffer layer 50 nm thick. On top of the other tunnel barrier
metal electrodes are deposited. The electron gas within the
electrodes is a source of the response potential which keeps the
electron wave function in a form of a stable wave packet
\cite{a3,a4} and sets it motion with a controlled trajectory by
application of voltages \cite{a1}. The electron motion in the growth
$y$ direction is frozen by the confinement within the quantum well.
We use the following Hamiltonian:
\begin{equation}
H(x,z,t)=-\frac{\hbar^2}{2m}\left(\frac{\partial^2}{\partial
x^2}+\frac{\partial^2}{\partial y^2}\right)-e\Phi(x,y_0,z,t)+H_D,
\end{equation}
where $y_0$ is the center of the quantum well, and $\Phi$ is the
electrostatic potential distribution inside the quantum well. This
potential is evaluated in a three dimensional region containing the
entire nanodevice by solution of the Poisson equation with the
method described in detail in Ref. \cite{a1,a2}. The calculation
method allows for description of the self-focusing effect \cite{a3}
and the time dependence of the potential distribution during motion
of the electron wave packet. The last term of the Hamiltonian (1)
introduces the spin-orbit coupling by the Dresselhaus term
\begin{equation}
H_D=\frac{\hbar k_{so}}{m}\left(p_x\sigma_x-p_z\sigma_z\right),
\end{equation}
where $\sigma_x$ and $\sigma_z$ are the Pauli matrices and the
characteristic wave vector $k_{so}$ depends on the quantum well
thickness, the electron band mass and the bulk coupling constant
$\gamma$:
$k_{so}=\frac{m}{\hbar}\left(\frac{\pi}{d}\right)^2\gamma$. The
state functions are written as two rows single column mattrices
\begin{equation}
\Psi(x,z,t)=\left(\begin{array}{c}\psi_1(x,z,t)\\\psi_2(x,z,t)\\
\end{array} \right).
\end{equation}
The simulations that we performed were based on the time dependent
Schr\"odinger equation \begin{equation}
\Psi(t+dt)=\Psi(t-dt)-\frac{2i}{\hbar}H(t)\Psi(t)dt.\end{equation}
For moving electron wave packet the electrostatic potential must be
evaluated in each time step which introduces the time dependence of
the Hamiltonian (1) entering into Eq. (4). In each simulation as
the initial condition we take a stationary eigenstate
\begin{equation}
H(x,z,0)\Psi(x,z,0)=E\Psi(x,z,0)
\end{equation}
with the potential distribution corresponding to a standing electron
packet.

\begin{figure}[ht!]
\centerline{\hbox{\epsfysize=100mm
               \epsfbox[67 355 522 820] {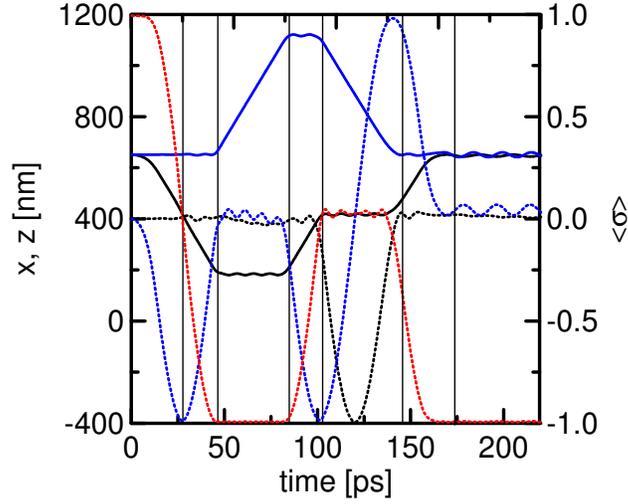}\hfill}}
\caption{Results of simulations for the NOT gate. Solid lines: black and blue
show the average values of $<x>$ and $<z>$ coordinates of the wave packet, respectively. Dotted lines:
black, blue and red show the average values of the Pauli matrix operators $<\sigma_x>$, $<\sigma_y>$
and $<\sigma_z>$. The solid vertical lines separate the time intervals
in which the spin is subject to subsequent operations:
$U_x(\phi)$,  $U_x(-\pi/2)$, $U_z(\pi)$, $U_x(\pi/2)$, $U_z(-\pi)$ and $U_x(-\phi)$.
\label{f3}}
\end{figure}

Below we apply ZnTe data parameters, in which the Dresselhaus
constant is equal to $\gamma=13.3$ eV {\AA}$^3$, the electron
effective mass $m=0.2$ and the dielectric constant $\epsilon=7.4$.
In this material the self-focusing effect is stronger than in GaAs
\cite{a3} due to larger effective mass and smaller dielectric
constant. The Dresselhaus coupling leads to rotation of the spin of
the electron following a straight line around the direction parallel
to the electron trajectory. The rotation angle depends on the
distance traveled by the electron and not on duration of the motion.
The rotation over a full angle occurs for the spin precession length
$\lambda_{so}=\frac{\pi}{k_{so}}$. Let us assume that the electron
runs in the $x$ direction and in time $t$ travels the distance
$w(t)$. Then, its spin is rotated around the $x$ axis by the angle
$\phi(t)=2\pi\frac{w(t)}{\lambda_{so}}$. The spin initially parallel
to the $z$ axis $(\phi(0)=0)$ after rotation by the $\phi$ angle
becomes parallel to
$\overrightarrow{\tau}(\phi)=\left(0,\sin(\phi),\cos(\phi)\right)$
vector. The spin projection operator on the $\overrightarrow{\tau}$
direction is expressed by the matrix operator
\begin{equation}
\sigma=\overrightarrow{\tau}\cdot
\overrightarrow{\sigma}=\left(\begin{array}{cc}
\cos(\phi)&-i\sin(\phi)\\ i\sin(\phi)&-\cos(\phi)\end{array}\right).
\end{equation}
The eigenequation for this operator
\begin{equation}
\sigma\chi_\pm=\pm\sigma_\pm
\end{equation}
is solved by the eigenvectors
\begin{equation}
\chi_+=\frac{1}{\sqrt{2}}\left(\begin{array}{c}\sqrt{1+\cos(\phi)}\\\frac{i\sin(\phi)}{\sqrt{1+\cos(\phi)}}\end{array}\right),\;\;
\chi_-=\frac{1}{\sqrt{2}}\left(\begin{array}{c}\frac{i\sin(\phi)}{\sqrt{1+\cos(\phi)}}\\\sqrt{1+\cos(\phi)}\\\end{array}\right).
\end{equation}
The operator transforming the basis of eigenstates of $\sigma_z$ to
the above basis has the form
\begin{equation} U_x(\phi)=\left(\begin{array}{cc}\sqrt{1+\cos(\phi)}&\frac{i\sin(\phi)}{\sqrt{1+\cos(\phi)}}\\\frac{i\sin(\phi)}{\sqrt{1+\cos(\phi)}}&\sqrt{1+\cos(\phi)}\end{array}\right).
\end{equation}
Similarly one obtains the operator of the spin rotation around the
$z$ axis for the electron moving along the $z$ axis
\begin{equation}
U_z(\phi)=\left(\begin{array}{cc}1&0\\0&\exp(i\phi)\end{array}\right).
\end{equation}
One can also derive the operator of the spin rotation around the $y$
axis
\begin{equation}
U_y(\phi)=\left(\begin{array}{cc}\sqrt{1+\cos(\phi)}&\frac{\sin(\phi)}{\sqrt{1+\cos(\phi)}}\\\frac{-\sin(\phi)}{\sqrt{1+\cos(\phi)}}&\sqrt{1+\cos(\phi)}\end{array}\right).
\end{equation}
In contrast to $U_z$ and $U_x$ the $U_y$ operator is not associated
with the time evolution of the spin, but it is still useful for the
system of reference transformations.

\begin{figure}[ht!]
\centerline{\hbox{\epsfysize=100mm
               \epsfbox[67 355 522 820] {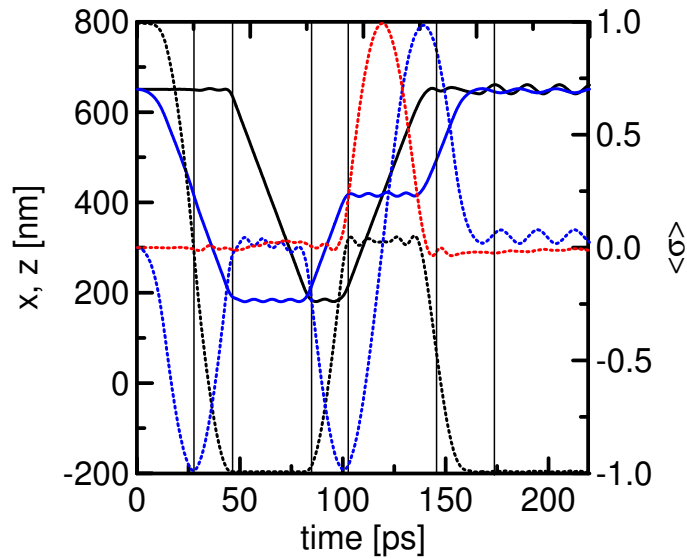}\hfill}}
\caption{Same as Fig. \ref{f3} but for the phase shift gate $U^\pi$. The solid vertical lines separate the time intervals
in which the spin is subject to subsequent operations:
$U_z(\phi)$,  $U_z(-\pi/2)$, $U_x(-\pi)$, $U_z(\pi/2)$, $U_x(\pi)$ and $U_z(-\phi)$.
\label{f4}}
\end{figure}
\section{Nanodevice}
In the device based on induced quantum wires the electron packet follows any desired trajectory and
consequently one can obtain an arbitrary spin rotation. For this purpose induced
dots and wires \cite{a1,a2} can be used. In Fig. \ref{f2} the system
of the electrodes on top of the structure is presented. In this
device the spin operation will be performed for small voltages
applied to the gates.  The system contains 13 electrodes serving to
various purposes. A small electrode $e_1$ plotted with the darkest
shade of grey localized in the center of the device generates a
quantum dot underneath it. The spin of the electron confined in this
quantum dot will store the quantum bit, on which the logical
operations will be performed. Pairs of electrodes ($e_2$, $e_3$),
($e_4$, $e_{10}$) and ($e_9$, $e_{12}$) or their parts which are
marked in Fig. 2 with medium grey shade are used to set the electron
in motion along closed loops. During the electron motion its spin
undergoes time evolution corresponding to the logical gates:
Hadamard, NOT and phase shift $\pi$ gate. Parts of electrodes $e_4$
and $e_9$ marked with lighter shade of grey serve to deliver the
electron from the quantum dot to the NOT or $\pi$ gates. The
electrodes plotted with the lightest grey are auxiliary and serve
only to symmetrize the induced potential in order to make the
electron trajectory as straight as possible.

\begin{figure}[ht!]
\centerline{\hbox{\epsfysize=100mm
               \epsfbox[67 355 522 820] {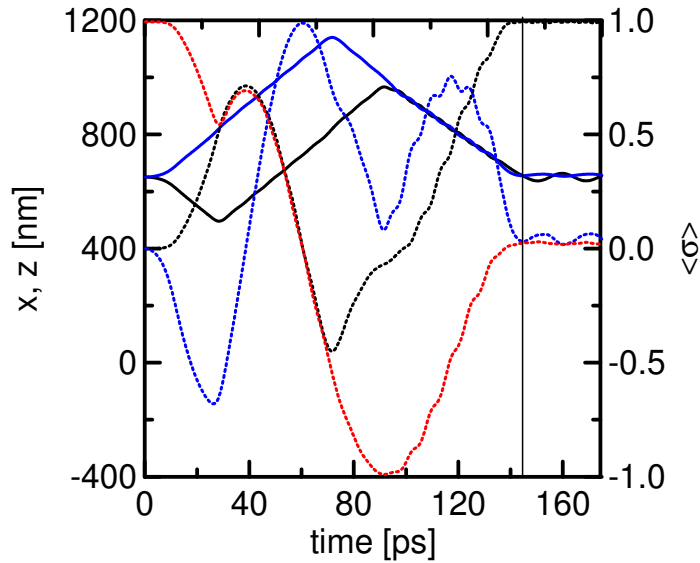}\hfill}}
\caption{Same as Figs. \ref{f3} and \ref{f4} but for the Hadamard gate $U^H$.
\label{f5}}
\end{figure}

The vertically oriented rectangle on the left part of the Figure
composed of the $e_{10}$ electrode and the darker part of $e_4$
electrode form the NOT gate. The electron going along the shorter
side passes a distance of $-\lambda_{so}/4$. Its spin is rotated
around the $x$ axis by $-\pi/2$ angle, the $U_x(-\pi/2)$ operation
is performed on the spin state. At the end of the segment parallel
to the $x$ axis there is a cut electrode corner \cite{a1}.
The electron packet is reflected of the cur corner under a right angle and
starts to move parallel to the $z$ axis on a distance of
$\lambda_{so}/2$ with the $U_z(\pi)$ operator acting on the spin
state. Then the electron moves along the $x$ and $-z$ directions
with $U_x(-\pi/2)$ and $U_z(-\pi)$ operators acting on the spin
state. Assembling all the spin rotations we obtain the NOT
transformation
\begin{eqnarray}
U&=&U_z(-\pi)U_x(\pi/2)U_z(\pi)U_x(-\pi/2)\\&=&
\frac{1}{\sqrt{2}}\left(\begin{array}{cc}1&-i\\-i&1\end{array}\right)\left(\begin{array}{cc}1&0\\0&-1\end{array}\right)\frac{1}{\sqrt{2}}\left(\begin{array}{cc}1&i\\i&1\end{array}\right)\left(\begin{array}{cc}1&0\\0&-1\end{array}\right)
\\&=&-i\left(\begin{array}{cc}0&1\\1&0\end{array}\right)=U^{NOT}
\end{eqnarray}
The electron is delivered to the NOT gate from under the electrode $e_1$
along the quantum wire induced under the lighter part of the
electrode $e_4$. We find that for the NOT gate the length of the
$e_4$ electrode is arbitrary, since the rotation of the spin along
is compensated by the one occurring for the electron going back to
the quantum dot
\begin{equation}
U_x(\phi)U^{NOT}U_x(-\phi)=U^{NOT}.
\end{equation}

\section{Results}
In the proposed device the operations on the electron spin are
induced by voltages applied to the electrodes. The subsequent steps
of the spin operations can be followed at Fig. \ref{f3} in which one
can observe the time dependence of the position of the center of
mass of the wave packet and the average values of the Pauli
matrices. The electron trajectory in the NOT gate is depicted in Fig. \ref{f2} by the black color. At the initial moment the same voltage of -0.2 mV is
applied to all the electrodes. The electron is localized under the $e_1$
electrode with the spin set (almost) parallel to the $z$ axis
($s_z\simeq +\hbar/2$). The NOT operation on the electron spin is
performed by application of zero voltage to $e_4$, $e_{10}$ and
$e_{11}$ electrodes. The electron is first attracted to under $e_4$
electrode, starts to move parallel to the $x$ axis and initially
increases its velocity until the entire packet is transferred under
$e_4$, when the velocity becomes constant and the motion of the
packet acquires a ballistic character. The electron goes to under
$e_{10}$ electrode from below the lighter part of $e_4$ electrode
and goes around the loop. At $t=100$ ps, when the electron is found
at the uppermost electrode $e_4$ part, we decrease the voltage on
$e_{10}$ and $e_{11}$ to -0.4 mV to make the electron reflect on the cut corner and leave the gate changing its
direction to parallel to the $x$ axis
in order to complete the full loop.
When the electron returns to
under $e_1$ electrode, with the voltage applied to this gate being
lower than the $e_4$ voltage, it looses its velocity and when it
stops at $t=175$ ps, we raise the voltage of $e_1$ to 0.1 mV to trap the
electron underneath. In Fig. \ref{f3} we notice oscillations of the
trapped electron position. They do not influence the electron spin
that acquires orientation opposite to the $z$ axis direction,
accordingly to the NOT gate operation.

The phase shift operation is performed in exactly analogical way for
the electron completing the loop along the rectangle of electrodes
placed at the bottom of the device. The entire operation on the
electron spin is performed by assembling separate rotations
\begin{equation}
U=U_z(-\phi)U_x(\pi)U_z(\pi/2)U_x(-\pi)U_z(-\pi/2)U_z(\phi)=i\left(\begin{array}{cc}1&0\\0&-1\end{array}\right)=U^{\pi}.
\end{equation}
The phase shift $\pi$ operation is performed for voltages $e_9$,
$e_{12}$ and $e_{13}$ set to zero at $t=0$. For $t=100$ ps the
voltage on electrodes $e_{12}$ and $e_{13}$ is lowered to $-0.4$ mV
to make the electron change its direction and to return to $e_1$
below $e_9$ electrode. Then for $t=175$ ps the electron is stopped
by the voltage on $e_1$ raised to 0.1 mV. The details of the time
evolution of the position and spin can be followed in Fig. \ref{f4}.
This time we deliver the electron to the gate along the electrode
parallel to the $z$ direction since for the $\pi$ gate the length of
the segment parallel to the $z$ direction is indifferent
$U_z(\phi)U^\pi U_z(-\phi)=U^\pi$.

The last rectangle of electrodes in Fig. \ref{f2} is rotated with
respect to the NOT gate by $\pi/4$ around the $y$ axis. This
rotation corresponds to the operator
\begin{equation}
U_y(\pi/4)=\frac{2^{-3/4}}{\sqrt{1}+\sqrt{2}}\left(\begin{array}{cc}1+\sqrt{2}&1\\-1&1+\sqrt{2}\end{array}\right).
\end{equation}
For the electron moving around the rotated structure the spin is
manipulated as
\begin{equation}
U=U_y^\dagger (\pi/4) U^{NOT} U_y(\pi/4)=\frac{1}{\sqrt{2}}
\left(\begin{array}{cc}1&1\\1&-1\end{array}\right)=U^H,
\end{equation}
in which we recognize the Hadamard transform. For the Hadamard
gate zero voltage is set to $e_3$. The electron is extracted to
under $e_3$ from under $e_1$. When the electron acquires a velocity
at $t=30$ ps we put zero the voltage on $e_2$ and after the electron
returns to under $e_1$ its voltage is raided to 0.1 mV. The results
of simulations are presented in Fig. \ref{f5}. Electron with spin
initially set parallel to the $z$ axis goes to the state in which it
is parallel to the $x$ axis, i.e. to an even superposition of states
parallel and antiparallel to the $z$ axis. In the proposed
nanodevice the Hadamard operation is accomplished in the shortest
time since $e_1$ is placed at the corner of the rectangle and the
electron does not need to be delivered from $e_1$ to the gate.

\section{Summary and Conclusion}
The self-focusing effect due to the interaction of the electron
packet with the charge induced on the electrodes placed on the
surface of the semiconductor nanostructure allows for the electron
transfer along a designed trajectory between chosen points in the
device in form of a wave packet of a stable shape. In a
semiconducting material in which the the spin-orbit coupling is
present the electron motion is accompanied by rotation of its spin
by the angle which depends on the direction of motion and the
distance traveled. For carefully chosen lengths of the segments one
can assembly the rotation into basic single qubit operations. For
electrodes forming closed loops with a common quantum dot at the
beginning and the end of each electron loop one can apply the
quantum gates sequentially in an arbitrary order. The nanodevice
proposed in the present paper is designed to perform any sequence of
operation of the most popular single qubit gates: NOT, Hadamard and
phase shift. The choice of the operation and its implementation
requires application of weak DC voltages. The performed computer
simulations were obtained by iterative solution of the time
dependent Schr\"odinger equation. The presented results contained
the time evolution of the position and spin of the electron and
their related modifications. The simulations indicate that the
operations designed algebraically can indeed be performed. For the
numerical calculations the material data of a specific material
(ZnTe). The proposed device has dimensions which are realistic and
can be produced by the present technology. Therefore, both the
processes of the spin rotations and the obtained operation times are
reliable.

\end{document}